# Periodic Concentration-Polarization-Based Formation of a Biomolecule Preconcentrate for Enhanced Biosensing

*Sinwook Park[1], Keren Buhnik-Rosenblau[2], Ramadan Abu-Rjal[1], Yechezkel Kashi[2] and Gilad Yossifon[1]\**

[1]Faculty of Mechanical Engineering, Micro- and Nanofluidics Laboratory, Technion – Israel Institute of Technology, Technion City 3200000, Israel

[2]Faculty of Biotechnology & Food Engineering, Technion – Israel Institute of Technology, Technion City 3200000, Israel

**ABSTRACT.** Ionic concentration-polarization (CP)-based biomolecule preconcentration is an established method for enhancing the detection sensitivity of target biomolecules. However, the formed preconcentrated biomolecule plug rapidly sweeps over the surface-immobilized antibodies, resulting in a short-term overlap between the capture agent and the analyte, and subsequently suboptimal binding. To overcome this, we designed a setup allowing for periodic formation of a preconcentrated biomolecule plug by activating the CP for predetermined on/off intervals. This work demonstrated the feasibility of the cyclic CP actuation and optimized the sweeping conditions required to obtain maximum retention time of a preconcentrated plug over a desired sensing region and enhanced detection sensitivity. The ability of this method to efficiently preconcentrate different analytes and to successfully increase immunoassay sensitivity, underscore its potential in immunoassays serving the clinical and food testing industries.

# Introduction

Highly sensitive detection of biomolecules, such as proteins, antigens, and antibodies, in biological samples, holds an enormous potential to satisfy the growing demands in various fields, including medical diagnostics, food safety, environmental monitoring and bio-defense applications.[1–4] One of classical approaches used for such detection is enzyme-linked immunosorbent assay (ELISA) within multiple-well plates, which relies on antibody-analyte binding, followed by quantification of the resulting complexes in the sample.[5] However, ELISA requires long incubation times and multiple washing steps, and large sample volumes. Thus, there is a considerable need for methods that significantly reduce detection time, demand smaller sample volumes and have higher detection sensitivity.[6]

Advances in microfluidic technology[7,8], have brought to the emergence of several strategies, such as steric filtration using membrane/nanogaps[9], isotachophoresis (ITP)[10–12], and ion concentration polarization (CP) [13–21], to produce on-chip preconcentrates of biomolecules. Such preconcentration accelerates reaction rates and improves detection sensitivity, while minimizing sample loss, which is particular value in samples with low-abundance biomolecules.[22,23]

Electrokinetically driven CP-based preconcentration is regarded as one of most efficient and common tools for enhancing detection of charged biomolecules in micro-scale bioanalysis.[14,24–26] Due to unique ion perm-selectivity of ion exchange membranes, imbalanced electromigration of ion flux occurs under non-equilibrium conditions, i.e. application of an external field, resulting in the formation of ion-depleted and ion-enriched layers at opposite membrane-electrolyte interfaces, called ion concentration-polarization (CP).[27–31] CP-based preconcentration[13,16,17,19,21,32–36], occurring at the outer edge of the depletion layer in an open microchannel-membrane system due to counteracting convective versus electromigrative fluxes of the third species, can continuously accumulate charged molecules of a sample solution and operate under a wide range of electrolyte conditions.[37] This method has been applied in immunoassays using either surface-patterned immobilized antibodies[33] or antibody-functionalized microbeads[22] to preconcentrate target biomolecules.

However, the main drawback of CP-based preconcentration systems is the inability to control the location of the preconcentrated biomolecule plug, and, with it, the inability to overlap the preconcentrated plug of target molecules and the surface-immobilized antibodies, a step required for enhanced detection sensitivity and binding kinetics. In cases where the antibodies are fixed to the surface of a microchannel, some precalibrations are necessary to ensure this overlap, which is very sensitive to the system parameters (e.g., flow rate, voltage, channel geometry). One means of achieving direct control over the length of the depletion layer, which, in turn, controls the location of the preconcentrated plug, uses either embedded electrodes or heaters for local stirring of the fluid, which induce alternating-current-electro-osmosis (ACEO)[38] and electro-thermal (ET) flow[39], correspondingly.

Here, we suggest a different approach, which induced periodic actuation of the CP in order to periodically sweep the preconcentrated plug of non-labeled (i.e. invisible) biomolecules over the target region. As a biosensing model, periodic CP was applied on human IgG as the target analyte detected using a standard sandwich ELISA format which used magnetic particle (MP)-conjugated capture antibodies trapped at the desired target area and a fluorescently-tagged detection antibody.

## Materials and Methods

### *Experimental setup*

The CP platform design was similar to previously studied open microchannel-Nafion interface devices.[40] The platform (Fig. 1a) was comprised of a polydimethysiloxane (PDMS) main microchannel (300-µm-wide, 12-mm-long, 40-µm-deep) with an embedded Nafion membrane (300-µm-wide, 1-mm-long) that interconnected the main microchannel and side chambers (3-mm-diameter). The two microchannel inlets were symmetrically electrically powered to minimize net electroosmotic flow (EOF) and to generate a stronger depletion region, while the side chambers were grounded, such that the microchannel behaved as the anodic side, wherein the depletion layer formed. For CP generation, four external platinum electrodes (0.5-mm-diameter) were inserted within each of the ends of the main channel and two side reservoirs and connected to a voltage source (Keithley 2636). Details of the chip wetting and cleaning steps prior to experiments are described elsewhere.[38,41]

To visualize the ionic concentration profiles, 5 $\mu$M pH-free Dylight molecules (Dylight 488, Thermo Scientific Inc.) were mixed in binding buffer (50mM Tris, 1% bovine serum albumin (BSA), 0.05% Tween 20; pH = 9.3, conductivity = 180 µS cm$^{-1}$). Green fluorescent protein (Aequorea victoria Green fluorescent protein/enhanced GFP, Sino Biological Inc.) was used at a concentration of 1 µg mL$^{-1}$. The net flow was driven by hydraulic pressure difference between microchannel inlets. The fluorescence intensity of the dye was analyzed by normalizing the local fluorescent dye intensity by that of the initial intensity measured before electric field application. All experiments were recorded with a spinning disk confocal system (Yokogawa CSU-X1) connected to an inverted microscope (Eclipse Ti-U, Nikon) and a camera (Andor iXon3).

*Binding of IgG and anti-IgG-conjugated particles*

Immunoassay targeting of human IgGs was conducted using a standard sandwich ELISA format, where the capture antibody was immobilized on the bead surface. The capture antibody (affiniPure mouse anti-human IgG (H+L), Jackson Immunoresearch Laboratories Inc.) was conjugated to 2.8 µm-diameter Dynabeads® magnetic beads (Rhenium) according to the manufacturer's instructions, and resuspended in binding buffer. Thereafter, 2 µL of the MP-conjugated ($5 \times 10^6$ bead/mL) were introduced into the microchannels and trapped at the desired locations (~ 2.3 mm upstream microchannel from the Nafion membrane interface) using an external magnet placed on the top surface of the PDMS. Then, 30 µL of serially diluted 1 µg/mL human IgG (ChromPure Human IgG, whole molecule, Jackson Immunoresearch Laboratories Inc.), were introduced and subjected to periodic CP during the reaction, as demonstrated in Fig. 1a (30 V, 20 min). After incubation, the analyte solutions were washed with binding buffer for 5 min by introducing a fresh buffer into the microchannel, and then replaced with binding buffer containing 0.5 µg/mL fluorescent detection antibodies (Alexa Fluor 488-conjugated AffiniPure Rabbit Anti-Human IgG; Jackson Immunoresearch Laboratories), which were allowed (10 min) to bind to the human IgG-MP-capture antibodies complex (step B in Fig 1a). The channel was then perfused with binding buffer for 5 min and the fluorescence intensities (area averaged of trapped magnetic particles after subtracting the background) of the bound detection antibodies were determined. In addition to standard sandwich ELISA format, an immunoassay with two components was performed, in which anti-human IgG antibody (Alexa Fluor 488-conjugated AffiniPure Rabbit Anti-Human IgG) and human IgG (ChromPure Human IgG)-conjugated to 2.8 µm-diameter MPs were used as the target analyte and the capturing probes respectively. Human IgGs were conjugated with the particles as per the manufacturer's instructions, and resuspended in binding buffer. All experiments were performed in the binding buffer in order to negatively charge the surface proteins of the human IgG.

*Numerical simulations*

To gain insight into the experimental results, they were qualitatively compared to a numerical solution. For simplification, instead of the three-dimensional (3D) structure of the microfluidic chip, we used the one-dimensional (1D) and one-sided setup of only one-half of the microchannel ($0 > \tilde{x} > \tilde{L}_c$, where $\tilde{L}_c$ is one-half of the channel length) interfacing an ideal permselective membrane [see Fig. 3(a)]. The mathematical model of the 1D, time-dependent ion transport in this setup in the dimensionless form reads (tilde notations are used below for the dimensional variables, as opposed to their untilded dimensionless counterparts):

$$\frac{\partial c_i}{\partial t} = -\frac{\partial j_i}{\partial x} = D_i \frac{\partial^2 c_i}{\partial x^2} + z_i D_i \frac{\partial}{\partial x}\left(c_i \frac{\partial \varphi}{\partial x}\right) - Pe \frac{\partial c_i}{\partial x}, \quad i = 1, 2, 3, \qquad (1)$$

$$2\delta^2 \frac{\partial^2 \varphi}{\partial x^2} = -\sum_{i=1}^{3} z_i c_i, \qquad (2)$$

where Eq. (1) is the Nernst-Planck equations satisfying the continuity of ionic fluxes. For convenience, we use $i=1$ for $K^+$, $i=2$ for $Cl^-$ and $i=3$ for negatively charged third species. $z_i$ and $D_i$ are the valence and the dimensionless diffusion coefficient (normalized by the diffusion coefficient of $K^+$, $\tilde{D}_1$) of species $i$, respectively, where $z_1 = -z_2 = 1$, $z_3 = -3$, $D_1 = D_2 = 1$ and $D_3 = 0.05$. All ion concentrations $\tilde{c}_i$ were normalized by the concentration of an electrically neutral solution (the outer stirred bulk concentration), $\tilde{c}_1 = \tilde{c}_2 - z_3 \tilde{c}_3 = \tilde{c}_0$. The spatial coordinate $\tilde{x}$ was normalized by one-quarter of the channel length $\tilde{L} = \tilde{L}_c/2$ (i.e. $0 < x < 2$), the time $\tilde{t}$ was normalized by $\tilde{L}^2/\tilde{D}_1$, and the ionic fluxes $j_i$ was normalized by $\tilde{D}_1 \tilde{c}_0/\tilde{L}$. Eq. (2) is the Poisson equation for the electric potential $\tilde{\varphi}$, which was normalized to the thermal potential $\tilde{R}\tilde{T}/\tilde{F}$, where $\tilde{R}$ is the universal gas constant, $\tilde{T}$ is the absolute temperature and $\tilde{F}$ is the Faraday constant. The symbol $\delta = \tilde{\lambda}_D/\tilde{L}$ is the Debye number, which is a small parameter equal to the ratio between the Debye length $\tilde{\lambda}_D = \sqrt{\tilde{\varepsilon}_0 \varepsilon_r \tilde{R}\tilde{T}/2\tilde{F}^2 \tilde{c}_0}$ and the characteristic scale of the problem. Herein, $\tilde{\varepsilon}_0$ and $\varepsilon_r$ are the permittivity of the vacuum and the relative permittivity, respectively. Finally, $Pe$ is the Peclet number and is defined as $Pe = \tilde{u}\tilde{L}/\tilde{D}_1$, where $\tilde{u}$ is the net flow speed.

At the membrane surface ($x = 0$), the following electrochemical boundary conditions were specified:

$$c_1(0,t) = N, \quad j_{2,3}(0,t) = 0, \quad \varphi(0,t) = -V - \ln N. \tag{3}$$

The first condition fixes the cation concentration in the ideally selective membrane at a value $N$ ($N > 0$ is the dimensionless fixed charge density of the membrane). The second condition imposes zero fluxes of the negatively charged ions transported across the membrane. The last condition sets the potential drop $V$ across the system ($-\ln N$ accounts for the Donnan potential jump). At the inlet, $x=2$ (the interface between the unstirred layer and the well-mixed bulk), the concentrations are fixed equal to the concentrations in the electrically neutral stirred bulk:

$$c_1(2,t) = 1, \quad c_2(2,t) = 1 + z_3 \bar{c}_{3,0}, \quad c_3(2,t) = \bar{c}_{3,0}, \quad \varphi(2,t) = 0, \tag{4}$$

where $\bar{c}_{3,0}$ is the concentration of the third species in the outer stirred bulk. The system (1)-(4) was solved numerically, taking equilibrium as an initial condition, for $\bar{c}_{3,0} = 10^{-3}$, $\delta = 10^{-3}$, $Pe = 210$, $V = 1600$ and $N = 25$ (Fig. 4).

## Results and Discussion

### *Sweeping the pre-concentrated plug by periodic CP application*

To examine the brushing (sweeping) effect of the preconcentrated plug over the target area, a periodic CP was applied by turning on/off the external voltages, as depicted in Fig.1b. The target area (714 µm × 286 µm), herein used as an interrogation window for the image analysis of the fluorescence intensity of the captured analyte (red rectangles in Fig 1b, c), was located upstream (2300µm) to the Nafion membrane edge. Under the operating conditions for the generation of CP ($u = 215 \pm 40$ µm s$^{-1}$ (Pe ~ 651), $V = 30$ V), the equilibrium location of the preconcentrated plug was formed further upstream (right) to the target area (red dashed rectangle) during a single CP cycle (Fig. 1c and 2). Figure 1c shows the periodic activation of the CP with time intervals of 65 s and 25 s for CP on and off, respectively. These intervals were experimentally chosen so as to ensure that the preconcentrated fluorescent molecules do not pass further upstream of the target area or downstream of the membrane during the on and off events, respectively. In the first interval of CP activation, $t_1$ (~120 s) was longer than the intervals of next periods due to initial generation of depletion. The preconcentrated plug of fluorescent molecules became intensified and wider with increasing time (Fig. 1b, c) due to the continuous accumulation of the molecules. These results clearly demonstrate that the location of the preconcentrated plug can be indirectly controlled by periodically turning the CP on/off.

*Optimal periodic CP conditions for increased overlap between the preconcentrated plug and the target area*

To understand the effect of periodic CP on concentration of the preconcentrated molecules over the target area, we applied various frequencies of periodic CP (i.e. time interval between on/off CP). The resulting sweeping behaviors of preconcentrated plug over the target area were evaluated by monitoring the bound fluorescent molecules (Dylight 488) or fluorescent proteins (GFP) as shown in Fig.2 and Fig.S1, respectively (see also movie S1). The experimental results were supported by numerical computations, depicted in Fig.3. Three frequencies of periodic CP (i.e., A, B, C) were selected in the experiments (see Fig. 2a) with time intervals of the off state $t_{\text{off\_A, B, and C}}$ of 50, 25 and 16 s, respectively. As a control, a constant voltage (no periodicity) was applied, resulting in the depletion layer passing over the target area.

Fig.2a A-D depicts the area-averaged intensity of the fluorescent molecules within the interrogation sensing area ($A$) described as $C = \frac{1}{A}\int_s C^* ds, \ C^* = C/C_{\text{initial}}$ as a function of time, with various frequencies of applied periodic CP. Note that the second peak in the periodic pattern of Fig.2a corresponds to the washout of the preconcentrated molecules during the off state as they pass over the target area. While for the periodic conditions depicted in part A, the peak of the fluorescent signal was the same over time, for the other cases of parts B-C, where the frequency of the CP application was increased, the magnitude of the peak increased with time (see also Fig.S1). The latter results from the continuous accumulation of fluorescent molecules in the oscillating plug, while in the former, the plug is completely flushed in between CP applications. The experimental results stand in qualitative agreement with the numerical simulations depicting the time-dependent concentration of a third species ($c_3$), normalized by its initial value (Fig. 3b). Qualitative agreement between experiments and numerical simulations was also obtained for the time integral of the fluorescent signal ($C_{\text{integral}} = \int_t \int_s \frac{1}{A} C^* ds \cdot dt$) over the sensing region for various applied CP frequencies. The results clearly demonstrated a greater overlap between the fluorescent molecules and the sensing region under periodic CP conditions as compared to non-periodic and no CP conditions. Increasing the frequency of the periodic CP (scheme B and C) results in an increase of the preconcentrated molecules over the target area. Such strategy of overlapping the target molecules and the sensing region seems to be applicable for molecules of a wide range of diffusion coefficients, e.g. Dylight 488[42] and GFP[43] fluorescently tagged molecules with ~430$\mu m^2 s^{-1}$ and 87$\mu m^2 s^{-1}$, respectively. Such a strategy is therefore useful for preconcentrating various types of target molecules over the desired sensing area.

*Enhanced binding of anti-human IgG to human IgG-conjugated magnetic particles using periodic CP*

As a proof of concept of periodic CP-driven increases in target biomolecule concentration, and to further demonstrate binding ability of biological assay components during CP induction, we preconcentrated a fluorescent anti-human IgG antibody over an area containing immobilized MP-conjugated human IgG. The use of fluorescently labeled analyte enabled visualization and tracking of the preconcentrated plug and also enabled determination of preservation of a fluorescent signal following a washing step, due to successful binding of the two immunological/biological components. Following the magnetic trapping of MP-conjugated human IgG, various concentrations of fluorescent anti-human IgG antibodies were introduced under forced flow (left to right direction, (73 ± 7 µm s$^{-1}$, Pe ~ 438)) in conjunction with periodic CP (30 V and time interval of 35s and 10s for on/off events, respectively) (see Fig.4a and Supplementary movie S2). Binding of the two immunological components with vs. without periodic CP was examined under various incubation times (Fig. S2). We chose an incubation time of 5 minutes for the tests shown in Fig.4 as it was the shortest time that we had tested (Fig.S2) but still exhibited a significant improvement in the binding performance relative to the no CP case. Results demonstrate that upon activation of periodic CP, the preconcentrated plug of anti-human IgG antibody molecules repeatedly swept over the trapped human IgG-MPs (Fig 4a), resulting in a significant improvement of the binding sensitivity, as assessed by fluorescence intensity after the washing step (Fig 4b, 4c). In particular, enhancement of the fluorescent signal was substantial at low anti-human IgG concentrations (5-fold enhancement at 0.25 µg/ml vs. ~1.5-fold enhancement at 2 µg/ml relative to the no CP case).

*Periodic CP-based sandwich immunoassay for detection of human IgG*

To model immunoassays typically used in clinical and food testing industries, we designed a sandwich ELISA format using a label-free (i.e., non-fluorescent) human IgG as the target analyte. MP-conjugated capture antibodies were immobilized in the target area, while fluorescent-anti human IgG antibodies served as detection antibodies. The operation parameters for the periodic CP were 30 V for the voltage amplitude, with time intervals of 100s and 30s for on/off event respectively, and a net flow of approximately 250 µm s$^{-1}$. Figure 5a shows the fluorescence images of the detection antibodies bound to the human MP-conjugated IgG-capturing antibody for different CP schemes, while Fig.5b depicts the intensity of the binding of human IgG in the sandwich immunoassay. A binding of human IgG was significantly enhanced upon application of periodic CP as compared to a single CP and no CP (Fig. 5b), which measured ten-fold higher signal intensity as compared to the no CP case, at 15 ng mL$^{-1}$ human IgG. In addition, testing of negative control chicken IgG clearly demonstrated the specificity of the detections of human IgG by the immobilized capture antibodies. The detection sensitivity of the system without CP activation (no CP) was similar to that of a routine standard sandwich ELISA performed in a multi-well plate (see Fig. S3). Apart from intensifying capture load, the microfluidic chip with periodic CP-based preconcentration required shorter incubation times and smaller sample volumes as compared to the conventional ELISA method.

## Conclusions

Sweeping a preconcentrated plug of biomolecules over a target region containing immobilized antibodies was realized by periodic application of CP. Invisible biomolecules (i.e., non-tagged antigens) can be preconcentrated periodically over the surface-immobilized antibodies without precise control of operating parameters (i.e., net flow rate, applied voltages). We determined the optimal sweeping conditions by determining the maximum retention time of the preconcentrated fluorescent plug over the target region, which showed good agreement with numerical simulations. To simulate a real biosensing application, a sandwich immunoassay was designed to trap human IgG in the CP-based platform and clearly demonstrated enhanced biosensing when using a periodic CP as compared to a single CP or no-CP at all. The presented tests proved the capability of the sandwich immunoassay to robustly preconcentrate target samples without requiring additional functions or pre-calibration steps. Optimization of the channel geometry, surface immobilization of capture antibodies, e.g., shifting to immobilization directly on the channel surface instead of using magnetic beads which can be sheared off and change their location under flow, is expected to further increase the detection sensitivity. The ability to efficiently preconcentrate different analytes and to successfully increase immunoassay sensitivity, emphasizes the potential of implementing periodic CP activation in immunoassays serving the clinical and food testing industries, especially when a highly sensitive method and a short analysis time are critical.

**AUTHOR INFORMATION**

**Corresponding Author**

*E mail: yossifon@technion.ac.il

**Author Contributions**

S.P, K. B.R, Y.K and G.Y planned the study. S.P fabricated the chips and performed, together with K.B.R, the experiments and analyzed the data. R.A.R performed the numerical simulations. G.Y and Y.K supervised the study. All authors contributed to preparation of the manuscript.

**Acknowledgements**

We acknowledge funding provided by the Israel Science Foundation (ISF), grant number 1938/16. Fabrication of the chip was made possible through the financial and technical support of the Russell Berrie Nanotechnology Institute and the Micro-nano Fabrication Unit.

# Figures

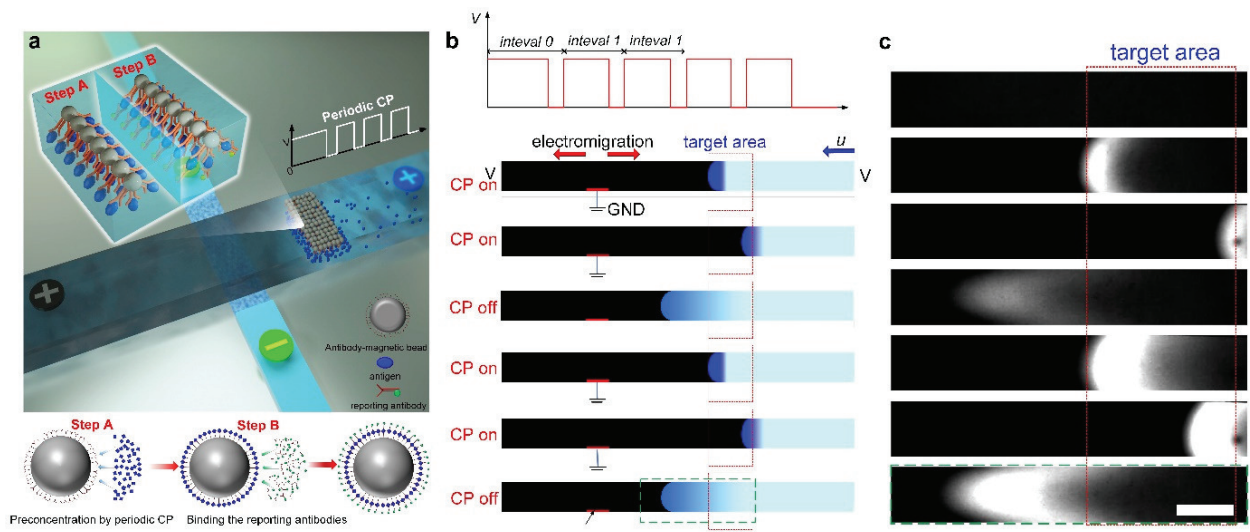

**Figure 1.** Periodic application of concentration-polarization and sweeping of the preconcentrated plug over the desired target area. (a) Schematics showing the periodic formation of a preconcentrated biomolecule plug, upon actuation of ionic concentration-polarization (CP), and the sandwich immunoassay approach wherein target biomolecules are bound between immobilized magnetic bead-conjugated antibodies and reporting antibodies. (b) Schematic illustration of periodic application of CP and the corresponding changes in location of the preconcentrated plug. (c) Time-lapse microscopy fluorescent images showing the periodic sweeping of the preconcentrated fluorescent molecules within the green-dashed rectangle depicted in (b). The times t0, t3 and t1, t4 correspond to the passage of the preconcentrated plug over the target area (red rectangles), while the times t2, t5 correspond to the release of the preconcentrated plug by deactivation of the CP. The applied voltage for CP activation was 30 V and red dot rectangle in a) and b) indicate the interrogation window defined as the target area. (scale bar: 300 μm).

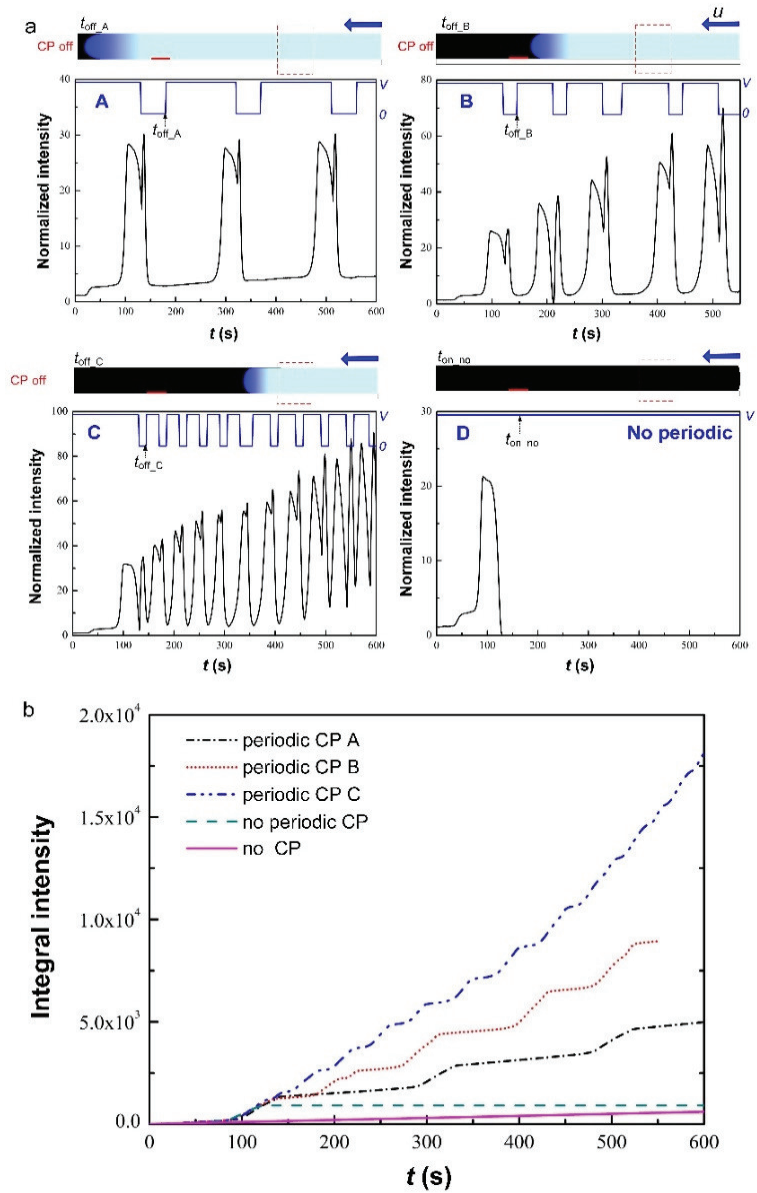

**Figure 2.** Transient concentration of fluorescent molecules within the sensing region and its dependence on the applied CP periodicity (see movie S1). a) Normalized intensity of the fluorescent molecules within the interrogation area as a function of time for various frequencies of the applied CP cycles. The periodically applied voltage (i.e., A, B, C) as well as the constant voltage applied (D) are indicated by the blue curves. The schematic insets indicate the momentary location of the released preconcentrated molecules before the next CP activation for periodic CP A, B, and C. In the case of constant voltage (no periodic CP application, D) the entire microchannel region was depleted. b) Time integral of retention of the fluorescent molecules over the target area.

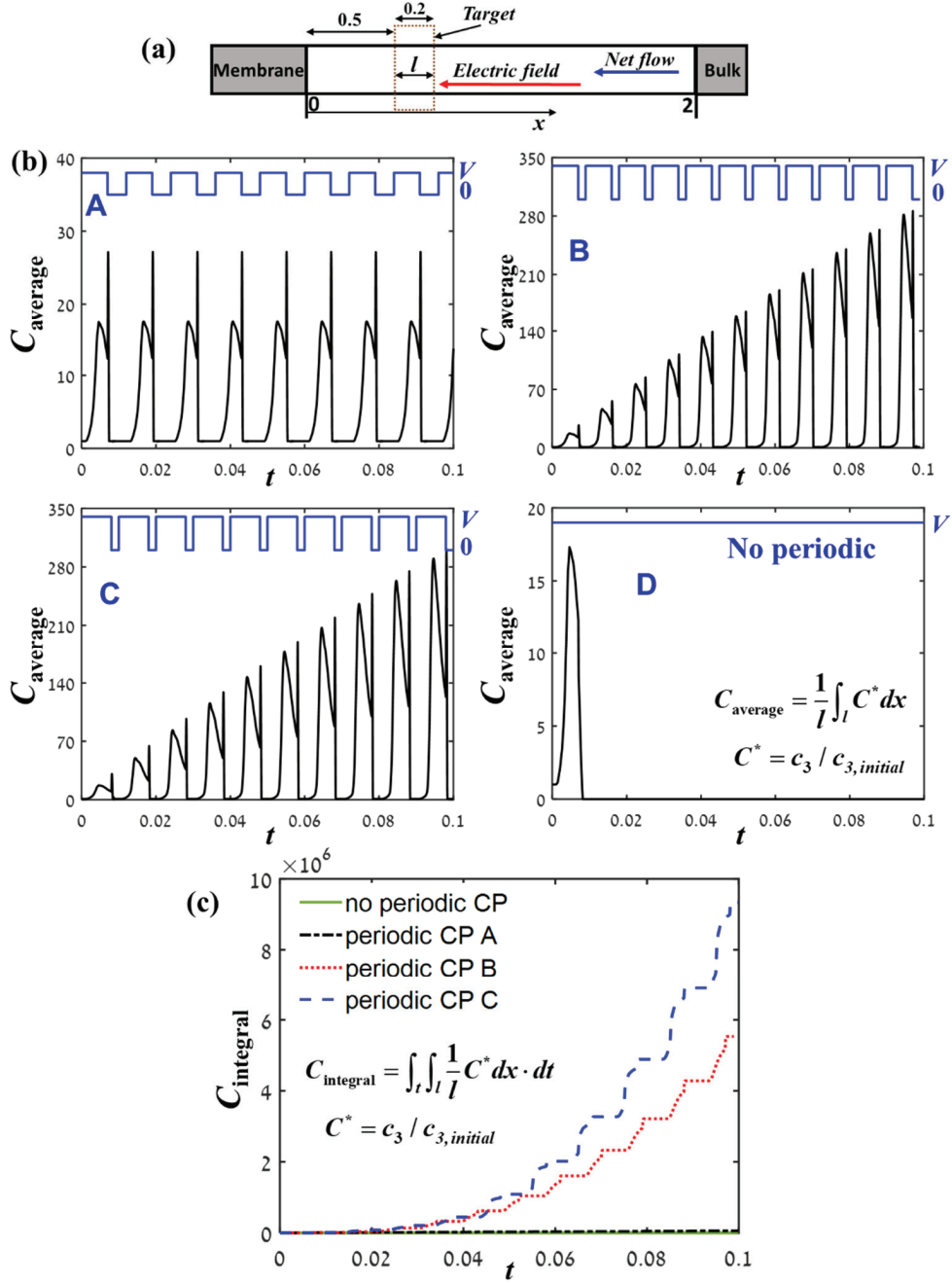

**Figure 3**. Numerical simulations of the transient concentration of a third species ($c_3$) within the sensing region, as a function of the periodic CP regime. a) Schematics of the 1D model. b) Normalized average concentrations of the third species within the interrogation segment for various frequencies of the applied CP. The periodic voltage applied (i.e., A, B, C) as well as the constant voltage (D) application are indicated by the blue curves. c) Time integral of the retention of the third species ions over the target segment.

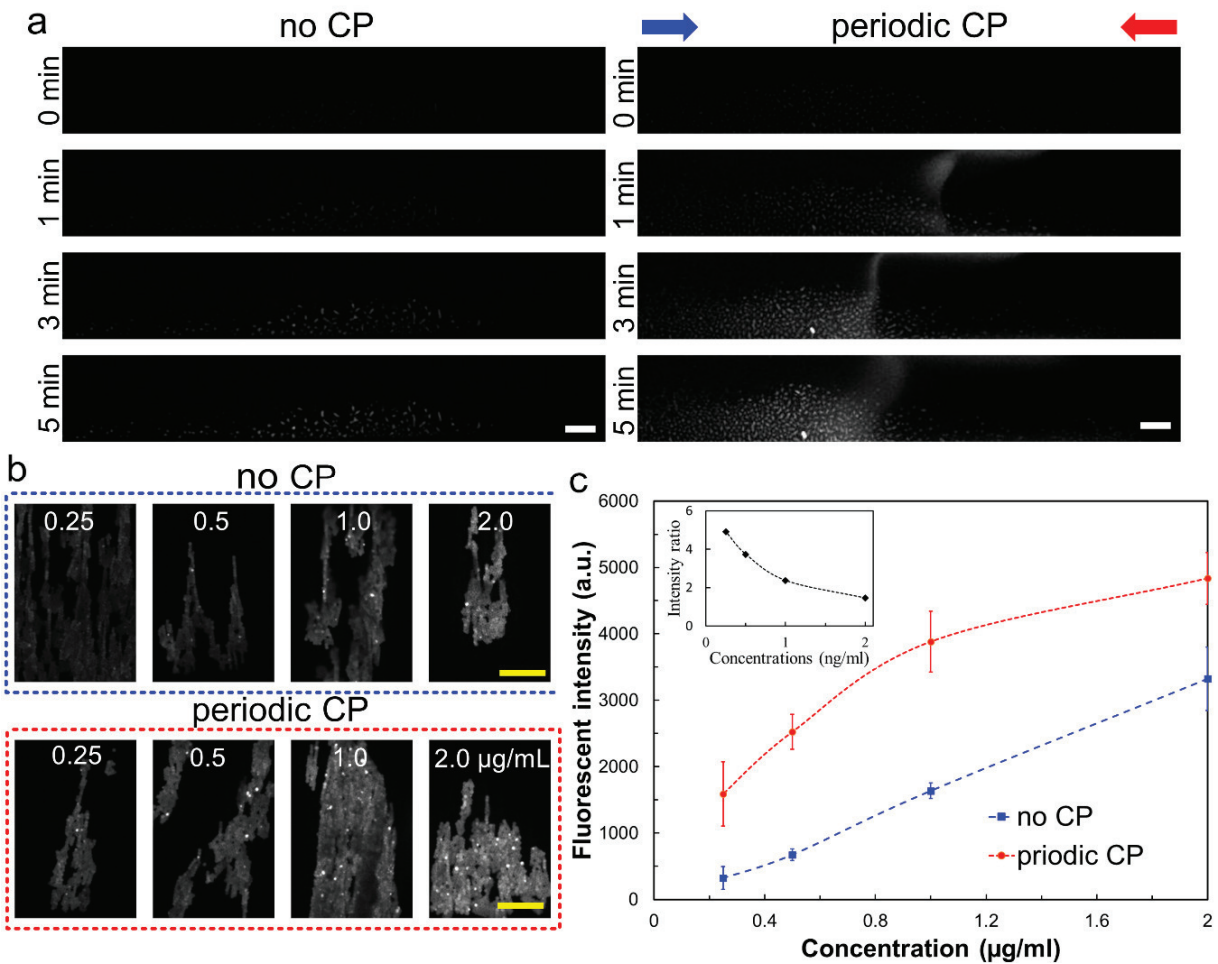

**Figure 4.** Enhanced biosensing efficiency upon application of periodic CP in a model composed of fluorescently labeled anti-human IgG antibody and magnetic particle-conjugated human IgG (human IgG-MPs). (a) Representative images showing the transient binding of fluorescently labelled anti-human IgG antibodies (1 µg ml-1) to human IgG-MPs, with versus without periodic CP activation. Microscope images (b) and corresponding fluorescence intensities measured following 5 minutes incubation and a washing step (c), as a function of the applied anti-human IgG concentration following. The inset indicates the fluorescence intensity ratio measured for samples projected to periodic CP (30 V) versus those without actuation of CP.

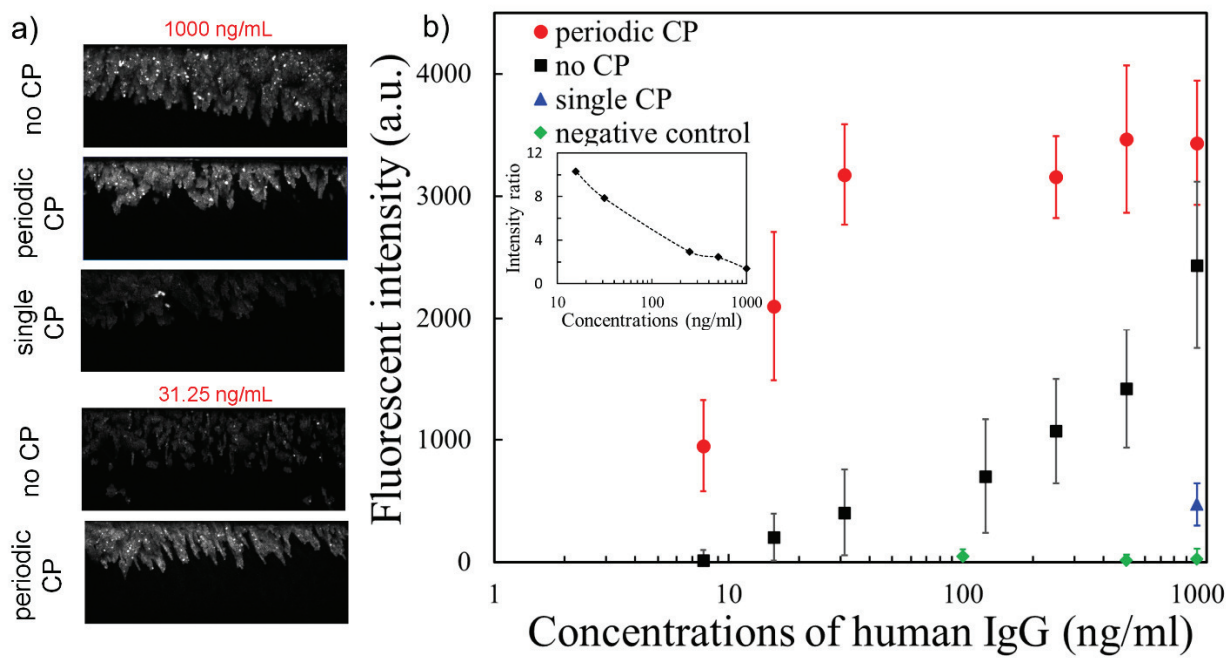

**Figure 5.** Enhanced biosensing upon application of periodic CP versus either a single CP or no-CP, in a sandwich immunoassay. a) Representative fluorescence microscope images showing the fluorescently tagged reporter antibody bound to magnetic particles in a sandwich immunoassay conducted under various operational regimes. b) Corresponding fluorescence intensities measured for various human IgG bulk concentrations. Chicken IgG, applied at various concentrations, served as a negative control. The inset indicates the fluorescence intensity ratio between periodic CP and no CP.